# Spin-canting-induced Giant Nonlinear Optical Magnetochirality in a 2D Ferrotoroid

Shian Xia[1,#], Sheng Liu[1,2,3,#,*], Wenhe Jia[4,#], Fanglu Qin[1], Xuanji Wang[1], Haixiang Luo[1], Yue Sun[1], Vanessa Li Zhang[1], Wenduo Chen[4], Jiazheng Qin[4], Cheng-Wei Qiu[4,5,*], Ting Yu[1,3,6,*]

[1]School of Physics and Technology, Wuhan University, Wuchang District, Hubei 430072, China.

[2]Wuhan National High Magnetic Field Center, Huazhong University of Science & Technology, Wuhan 430074, China.

[3]Wuhan Institute of Quantum Technology, Wuhan 430206, China.

[4]Department of Electrical and Computer Engineering, National University of Singapore, Singapore 117583, Singapore.

[5]National University of Singapore Suzhou Research Institute, Suzhou 215000, China.

[6]Key Laboratory of Artificial Micro- and Nano-structures of Ministry of Education, Wuhan University, Wuhan 430072, China.

# These authors contributed equally to this work.

* Correspondence to E-mail: yu.ting@whu.edu.cn; chengwei.qiu@nus.edu.sg; liu.sheng@whu.edu.cn

## Abstract

Achieving magnetically switchable chiral light emission is an important goal for 2D opto-spintronics. However, conventional strategies face a fundamental trade-off between dynamic tunability and polarization contrast. Nonlinear optics, particularly the emerging mechanism of chiral second-harmonic generation (SHG), offers a distinct strategy to bypass this restriction, yet its experimental realization remains elusive due to stringent symmetry requirements. Here, we report giant nonlinear optical magnetochirality in a centrosymmetric 2D ferrotoroid, bilayer (2L) CrSBr. We reveal that a field-induced spin-canting state breaks the parity-time ($\mathcal{PT}$) symmetry of the unperturbed antiferromagnetic (AFM) ground state, activating a spin-chirality-driven $i$-type susceptibility. The coherent interference between this emergent $i$-type and intrinsic $c$-type SHG susceptibilities generates a macroscopic circularly polarized SHG signal whose helicity is magnetically switchable. Leveraging this sensitive mechanism, we uncover remanent magnetic states after field saturation that evade conventional linear probes. By exploiting the non-volatility of these states, we demonstrate magneto-optical memory and logic operations. Our work establishes a general symmetry-driven strategy for tailoring nonlinear magnetochirality, while providing a sensitive optical probe for subtle spin textures in the 2D limit.

## Introduction

Magneto-optical spectroscopy has emerged as an indispensable probe for resolving microscopic magnetic states and textures, and their coupling to photonic degrees of freedom in two-dimensional (2D) quantum materials[1-3]. Recently, this field has witnessed transformative breakthroughs, ranging from unveiling intrinsic 2D ferromagnetism[4-6] and compensated antiferromagnetic (AFM) orders[7-11], to manipulating valley-spin locking[12-15] and coherent light-spin coupling interfaces[16-19]. While these capabilities are profound, a pivotal challenge yet to be conquered is the dynamic, reversible control over chiral light emission using magnetic states. Such control underpins next-generation opto-spintronic applications, spanning from high-capacity optical communication to magnetically controlled all-optical logic computing[20-22]. To achieve this, researchers have leveraged either intrinsic degrees of freedom such as spin-polarized injection[23,24] and valley-polarized excitons[25-27], or artificial architectures like plasmonic metasurfaces[28-31] and chiral molecular assemblies[32,33]. However, a fundamental trade-off persists: structurally fixed platforms lack dynamic reconfigurability, whereas existing tunable systems suffer from poor polarization contrast, leaving high-contrast, magnetically switchable chiral emission a long-standing challenge.

Recently, a distinct nonlinear pathway termed chiral second-harmonic generation (SHG) has been proposed in magnetic materials[34], departing from the conventional dichotomy where magnetic and crystallographic responses are strictly assigned to $c$-type and $i$-type susceptibilities, respectively[35]. Despite its recent emergence, its potential for realizing magnetically switchable chiral emission remains unexplored. Here, a novel $i$-type susceptibility is activated by vector spin chirality under broken parity-time ($\mathcal{PT}$) symmetry[34,36,37]. In contrast, the $c$-type susceptibility is governed by the primary magnetic order parameter, arising from the spatial inversion ($\mathcal{P}$) symmetry breaking induced by broken time-reversal ($\mathcal{T}$) symmetry. Driven by deterministic magnetic configurations, both components can thus be manipulated simultaneously. Strikingly, in the ideal non-dissipative limit, symmetry dictates that the $\mathcal{T}$-odd $c$-type and $\mathcal{T}$-even $i$-type tensors are purely imaginary and purely real, respectively[38]. This intrinsic phase orthogonality drives a coherent interference between the two channels, providing a deterministic avenue to translate microscopic magnetic orders directly into macroscopic, switchable photon helicity.

Despite its theoretical promise, the experimental realization of nonlinear optical magnetochirality remains elusive, as it inherently demands the simultaneous fulfillment of three stringent conditions at the level of the nonlinear optical response. First, the material must possess a centrosymmetric lattice. This eliminates the conventional crystallographic $i$-type SHG background, which has historically been the primary obstacle masking genuine chiral SHG effects[35,39]. Second, to synthesize macroscopic circularly polarized emission, the $c$-type ($\mathcal{T}$-odd) and $i$-type ($\mathcal{T}$-even) susceptibilities must populate orthogonal tensor elements, ensuring that their interference provides the required transverse phase retardation. Finally, the magnetic point group must be susceptible to external magnetic fields, enabling a deterministic symmetry transition from a $\mathcal{PT}$-preserved state to a $\mathcal{PT}$-broken state. This dynamic symmetry breaking acts as a physical switch, toggling the chiral $i$-type SHG from zero to a finite magnitude.

Here, we report the spin-canting-induced giant nonlinear optical magnetochirality in a centrosymmetric 2D ferrotoroid, bilayer (2L) CrSBr. We reveal that in its unperturbed AFM ground state, 2L CrSBr strictly preserves $\mathcal{PT}$ symmetry and permits only linearly polarized $c$-type SHG. However, a field-induced spin-canting state dynamically breaks the $\mathcal{PT}$ symmetry, activating a spin-chirality-driven $i$-type susceptibility. The coherent interference between this emergent $i$-type and the intrinsic $c$-type SHG susceptibilities generates a macroscopic circularly polarized SHG signal with magnetically switchable helicity. Leveraging the sensitivity of this interference mechanism, we uncover remanent magnetic states that evade conventional linear probes following field saturation. Exploiting the non-volatility of these states, we demonstrate a magneto-optical encoding and a high-contrast magneto-optical XNOR operation. Our work establishes a general symmetry-driven strategy for tailoring nonlinear magnetochirality, while providing a sensitive optical probe for subtle spin textures in the 2D limit.

# Results

## Symmetry design for switchable magnetochirality

To achieve switchable optical magnetochirality in 2D materials, we propose a symmetry-driven design strategy based on the coherent interference of nonlinear optical responses driven by distinct macroscopic order parameters. Generally, electric-dipole second-harmonic generation (ED-SHG) requires the breaking of $\mathcal{P}$ symmetry, which in magnetic materials originates from three distinct mechanisms (Fig. 1a). Conventionally, $\mathcal{P}$ symmetry breaking by the crystal lattice leads to a crystallographic $i$-type SHG, whereas its breaking by $\mathcal{T}$-broken magnetic ordering yields a $c$-type SHG. Crucially, non-collinear or canted spin textures can break both $\mathcal{P}$ and $\mathcal{PT}$ symmetries, activating a spin-chirality-driven $i$-type SHG (chiral SHG). In contrast to the crystallographic $i$-type SHG, both the $c$-type and chiral SHG are sensitive to the underlying magnetic order, enabling dynamic modulation via external magnetic fields. However, isolating these magnetic contributions is typically obscured by the crystallographic background. To realize optical magnetochirality, our strategy relies on a centrosymmetric magnetic system to eliminate this crystallographic $i$-type background. Here, we demonstrate a symmetry-driven strategy exploiting the mechanism of switchable magnetochirality via field-induced $\mathcal{PT}$ symmetry breaking (Fig. 1b). In the collinear $\mathcal{T}$-broken ground state where $\mathcal{PT}$ symmetry is preserved, the $i$-type component is forbidden, leaving only the intrinsic $c$-type SHG that manifests as linearly polarized emission. The application of an external magnetic field induces a spin-canting state, which breaks the $\mathcal{PT}$ symmetry and activates the $i$-type susceptibility alongside the intrinsic $c$-type contribution. Crucially, in the ideal non-resonant limit, the anti-unitary nature of the time-reversal operator dictates that $\chi^{(c)}$ and $\chi^{(i)}$ are imaginary and real, respectively[38]. This orthogonal phase relationship mathematically ensures that their coherent superposition yields non-vanishing helicity-dependent interference in the emitted SHG, generating a macroscopic chiral response. Specifically, for an AFM ground state, a perpendicular magnetic field leaves the Néel vector invariant but reverses the vector spin chirality, thereby flipping the sign of the chirality-induced $\chi^{(i)}$ while leaving the intrinsic $\chi^{(c)}$ unaffected. This selective sign reversal toggles the chiral interference, generating switchable left- ($\sigma^+$) or right-handed ($\sigma^-$) circularly polarized emission.

From a material-design perspective, these optical requirements translate into three magnetic-symmetry conditions. First, the crystal lattice must possess centrosymmetry ($\mathcal{P}_{lattice}$ symmetry) to eliminate the crystallographic $i$-type background. Second, the magnetic ground state must break $\mathcal{T}$ symmetry to accommodate the intrinsic $c$-type SHG. Third, the spin texture must couple with an external magnetic field to break the $\mathcal{PT}$ symmetry, activating the chiral SHG. As illustrated in Fig. 1c, while conventional magnetic materials satisfy only a subset of these prerequisites[4,6-10,35,39-48], spin-canted 2L CrSBr occupies their intersection[11,34,49,50]. While a comprehensive group-theoretical classification confirms the extreme rarity of this exact symmetry intersection[45], 2L CrSBr perfectly satisfies all three conditions. As detailed by its magnetic point group evolution under varying external fields (Supplementary Table S1), this metastable canted system provides a platform to realize and manipulate macroscopic optical magnetochirality.

**Intrinsic $c$-type SHG in the antiferromagnetic ground state**

To verify the symmetry prerequisites, we first characterize the $c$-type SHG of 2L CrSBr in its zero-field ground state. At zero magnetic field, 2L CrSBr exhibits an A-type AFM order, where the intralayer spins are ferromagnetically aligned along the crystallographic $b$-axis and couple antiferromagnetically across the vdW gap (Fig. 2a). While the crystal lattice is centrosymmetric (point group $mmm$), the layered AFM spin configuration breaks both $\mathcal{P}$ and $\mathcal{T}$ symmetries while preserving the combined $\mathcal{PT}$ symmetry. This specific magnetic symmetry establishes 2L CrSBr as a 2D ferrotoroid[51,52], intrinsically allowing $c$-type SHG while forbidding $i$-type background. Confirming its second-order nonlinear nature, the SHG intensity scales quadratically with the fundamental excitation power, exhibiting a logarithmic slope of ~1.938 (Fig. 2b). Its magnetic origin is confirmed by the temperature dependence (Fig. 2c): the SHG signal emerges below the Néel temperature ($T_N \approx 135.1$ K) and follows the critical power law $I(2\omega) \propto (T_N - T)^{2\beta}$. The extracted exponent of $2\beta \approx 0.45$ is consistent with the theoretical value ($2\beta = 0.46$) for the 2D XY universality class[53].

Crucially, the SHG response is governed by the evolution of its magnetic point group under an external field. While the AFM ground state (magnetic point group $m'mm$) allows the $c$-type SHG, applying a saturation magnetic field along the $b$-axis fully polarizes the spins into a ferromagnetic (FM) state (Fig. 2d). This spin-flip transition alters the magnetic point group to $m'mm'$, restoring the $\mathcal{P}$ symmetry and forbidding all ED-SHG processes (Supplementary Note 1). Experimentally, we track this process by monitoring both the exciton photoluminescence (PL) and the SHG intensity (Fig. 2e, 2f). Driven by the strong spin-layer coupling, the PL energy exhibits a characteristic step-like redshift upon the transition to the FM state, serving as an established signature[54]. Concurrently, the SHG intensity quenches to zero at ~0.15 T, and its hysteretic window coincides with the PL shift. This SHG on/off switching verifies the magnetic origin of the zero-field signal, confirming it as an intrinsic $c$-type SHG free from structural background.

**Giant magnetically switchable chirality via $\mathcal{PT}$ symmetry breaking**

Having established the intrinsic $c$-type SHG in the collinear AFM ground state, we next investigate the magnetochirality induced by an out-of-plane external magnetic field. We performed circularly-resolved SHG spectroscopy to map the evolution of the emission helicity. Here, we use the term SHG circular dichroism (CD) to denote the emitted-helicity intensity difference, $\Delta I_{\text{CD}} \equiv I_{\sigma^+} - I_{\sigma^-}$. As shown in Fig. 3a, at zero magnetic field, the emission intensities for left- $(\sigma^+)$ and right-handed $(\sigma^-)$ circular polarizations are nearly identical. This vanishing SHG CD is consistent with the $\mathcal{PT}$-preserved $m'mm$ symmetry of the collinear AFM state, which forbids the time-invariant $i$-type susceptibility. In contrast, upon applying an out-of-plane magnetic field of +1.2 T, the SHG emission becomes almost exclusively left-handed $(\sigma^+)$, yielding a giant SHG CD with a DOCP of approximately 81% (Fig. 3b). Upon reversing the magnetic field to -1.2 T, the dominant emission helicity switches to right-handed, yielding a reversed DOCP of ~ -77% (Fig. 3c). This deterministic switching highlights the reversible control over the nonlinear optical magnetochirality.

Fundamentally, the macroscopic effect originates from a field-induced symmetry reduction. As illustrated in Fig. 3d, the out-of-plane magnetic field cants the spins by an angle $\theta$, reducing the magnetic point group from $m'mm$ to $m'2'm$ (Supplementary Table S1 and Note 1). The consequent breaking of $\mathcal{PT}$ symmetry activates the $i$-type susceptibility ($\chi^{(i)}$). As established earlier, this emergent $i$-type susceptibility interferes coherently with the intrinsic $c$-type susceptibility ($\chi^{(c)}$). To validate the coherent interference mechanism, we map the continuous evolution of the SHG CD ($\Delta I_{CD}^{2\omega}$) as a function of the out-of-plane magnetic field (Fig. 3e). The experimental data are reproduced by the derived theoretical scaling law: $\Delta I_{CD}^{2\omega} \propto \mathrm{Im}\left[\chi^{(c)} \cdot \chi^{(i)*}\right] \propto sin2\theta \cdot cos\theta$(Supplementary Note 2). The quantitative agreement confirms that the SHG CD originates from the coherent cross-terms between the intrinsic AFM Néel vector $\boldsymbol{L} = \boldsymbol{S_1} - \boldsymbol{S_2}, \left(\chi^{(c)} \propto L \propto cos\,\theta\right)$ and the field-induced vector spin chirality $\boldsymbol{\kappa} = \boldsymbol{S_1} \times \boldsymbol{S_2}, \left(\chi^{(i)} \propto \kappa \propto sin(2\theta)\right)$. Furthermore, temperature-dependent circularly resolved SHG spectroscopy underscores the robust thermal stability of this nonlinear optical magnetochirality (Fig. S1). The DOCP remains above 70% up to 90 K and reaches approximately 90% near 30 K. This broad operational temperature window highlights the practical viability of exploiting spin-chirality-driven nonlinear phenomena for liquid-nitrogen-temperature opto-spintronic applications.

**Non-volatile memory and XNOR logic encoded by remanent spin chirality**

Having established the coherent interference mechanism for achieving the giant magnetochirality, we next extend this framework to linearly polarized rotational anisotropy (RA) SHG. To explore the sensitivity of this symmetry-selective probe, we measured the RA-SHG patterns at 0 T following magnetic saturation at +2 T and -2 T. Theoretically, a collinear AFM ground state (governed only by $\chi^{(c)}$) yields a symmetric four-lobe RA-SHG pattern under parallel (XX) polarization (Supplementary Note 3 and Fig. S2). However, the measured patterns exhibit a geometric distortion featuring two major and two minor lobes, deviating from the response expected for an ideal collinear AFM state. Instead, the system relaxes into two macroscopic states depending on the magnetic history (Fig. 4a, 4b). The principal polarization axes (major lobes, labeled $P^+$ and $P^-$) rotate in opposite directions, yielding absolute angles of ~35° and 145°, respectively. As derived in Supplementary Note 3, this symmetry-broken rotation is the macroscopic optical signature of a remanent $\chi^{(i)}$, which we attribute to a minute spin canting. In 2L CrSBr, this metastable canted state is trapped by the competition between interlayer exchange interactions and intrinsic magnetic anisotropy[55-57]. Theoretically, the energetic frustration creates a shallow energy landscape. At low temperatures, local ferromagnetic interactions effectively act as pinning centers within these minima, trapping the system in this metastable canting configuration. As this remanent canting angle is minute, it eludes conventional linear optical probes like PL measurements (Fig. S3). In contrast, chiral $i$-type SHG, serving as a symmetry-selective zero-background probe, resolves this trace magnetic signature. To confirm that this metastable memory effect is a global phenomenon, we performed spatial SHG intensity mapping projected onto the $P^+$ and $P^-$ polarization axes (Fig. 4c). For the positive remanent state (+2 T → 0 T), the entire flake is homogeneously dominated by the $P^+$ component. In contrast, this intensity distribution reverses ($P^- > P^+$) for the negative remanent state (-2 T → 0 T). Furthermore, we evaluated the long-term data retention of these remanent states. By continuously tracking the Degree of Linear Polarization ($DOLP = \frac{I(P^+)-I(P^-)}{I(P^+)+I(P^-)}$), we observe a non-decaying plateau over 48 hours (Fig. 4d), demonstrating the long-term stability of the remanent magnetic state.

Leveraging this non-volatility, we use the DOLP as the binary readout signal to encode the remanent magnetic states. The binary states are magnetically ‘written’ by applying and subsequently removing out-of-plane field of opposite polarities and are non-destructively ‘read’ through the sign of the DOLP, which switches deterministically and reversibly between positive and negative values (‘1’ and ‘0’) (Fig. 4e). Building upon this read-write functionality, we demonstrate a magneto-optical XNOR logic operation by using the incident linear polarization and the remanent magnetic state as two binary inputs. Specifically, by assigning the linear polarization axes ($P^{+}$ and $P^{-}$) and the opposite magnetic polarities as two-state logic inputs, the read-out SHG intensity deterministically maps out the XNOR truth table, where the strong and weak signals naturally act as ‘high’ and ‘low’ logic levels (‘1’ and ‘0’) (Fig. 4f). This integration of non-volatile memory and logic functionalities highlights the potential of remanent spin states for magneto-optical information processing.

## Conclusions

In summary, we have demonstrated a giant, magnetically switchable nonlinear chiral optical response in a centrosymmetric 2D magnet, driven by $\mathcal{PT}$ symmetry breaking. By establishing the coherent interference between the intrinsic $c$-type and the field-activated chiral $i$-type susceptibilities, we mapped the microscopic spin-canting configurations directly onto the macroscopic helicity of the SHG photons. Furthermore, this deterministic magneto-chiral interference mechanism enables the non-destructive optical readout of remanent spin chirality, providing a robust physical foundation for non-volatile optical memory.

# Methods

## CrSBr single crystal growth

CrSBr single crystals were synthesized via chemical vapor transport (CVT, Wuhan Shiwei Optoelectronic Technology Co., Ltd.). Chromium (III) bromide ($CrBr_3$, 99.00% purity, Macklin), sulfur powder (S, 99.95% purity, Aladdin), and Chromium powder (Cr, 99.00% purity, thermo scientific) were used as precursors. A stoichiometric mixture of $CrBr_3$ (720 mg), S (196 mg), and Cr (189 mg) was sealed within an evacuated quartz ampoule using an oil-free molecular pump under a high vacuum. The ampoule was subjected to a controlled thermal gradient. The source zone temperature was programmed as follows: ramped to 500 °C over 6 h, then to 550 °C over 12 h, subsequently to 930 °C over 12 h, and finally held at 930 °C for 84 h before cooling naturally to room temperature. Concurrently, the growth zone temperature profile was: ramped to 570 °C over 6 h and held for 12 h, then increased to 970 °C over 12 h and maintained for 24 h, followed by cooling to 850 °C over 12 h and holding for 48 h, before natural cooling to room temperature. The resulting crystals were recovered and stored within an argon-filled glove box upon opening the ampoule. In our experiment, 2L CrSBr flakes were mechanically exfoliated onto a freshly cleaned $SiO_2$/Si substrate. The precise 2L thickness was identified by its characteristic PL exciton energy and peaks.

## SHG and PL measurements

The optical SHG and PL measurements were performed using a Witec Alpha 300R confocal microscope integrated with a closed-cycle helium cryostat (attoDRY2100, attocube). The system provides a stable helium environment with base temperatures down to 1.8 K and is equipped with a superconducting magnet capable of applying magnetic fields up to 9 T (out-of-plane) and 3 T (in-plane). A 1064 nm linearly polarized pulsed laser was focused perpendicular to the sample surface through a 69.94× objective lens (numerical aperture = 0.82). The reflected signals, including the SHG emission and the exciton PL, were collected by a photonic crystal fiber and coupled into a spectrometer equipped with a 150 g/mm grating. For circularly-resolved SHG measurements, spectra were acquired for all four excitation–collection helicity combinations. The emitted-helicity-resolved intensities were calculated as $I_{\sigma^+} = I_{LL} + I_{RL}$ and $I_{\sigma^-} = I_{LR} + I_{RR}$. The degree of circular polarization was then calculated as $\mathrm{DOCP} = \frac{I_{\sigma^+} - I_{\sigma^-}}{I_{\sigma^+} + I_{\sigma^-}}$. For RA-SHG measurements, the polarization-resolved configurations were obtained by rotating the half-wave plate in the analyzer. For the circularly-resolved SHG measurements, quarter-wave plates were inserted to modulate and analyze the left- and right-handed circular polarizations. All measurements were conducted under a low-temperature helium environment unless otherwise specified.

## Data Availability

The data that support the findings of this study are available from the corresponding author upon reasonable request.

## Acknowledgement

We acknowledge the assistance from the National Key Research and Development Program of China (No.2021YFA1200800 to T.Y.), the National Natural Science Foundation of China (Grant Nos. 12574129 to S.L.), the Start-up Funds of Wuhan University to T.Y. and the Open Research Fund of the Pulsed High Magnetic Field Facility (Grant No. WHMFC2024017 to S.L.). C.-W.Q. acknowledges the financial support by the Ministry of Education, Republic of Singapore (Grant No.: A-8002152-00-00 & A-8004709-00-00), and the Competitive Research Program Award (NRF-CRP26-2021-0004 & NRF-CRP30-2023-0003) from the National Research Foundation, Prime Minister's Office, Singapore. This work is supported by the Advanced Research and Technology Innovation Centre (ARTIC), the National University of Singapore under Grant (project number: AFP-RP4). This work is also supported by National Key Research and Development Program of China (2023YFF0613600).

## Author Contributions

S.X. conceived and designed the experiments. X.W. grew the single crystals. S.X. and F.Q. fabricated the samples. S.X., Y.S., and H.L. carried out the SHG and PL measurements. S.X., W.J., W.C., and J.Q. analyzed the experimental data. S.L., C.-W.Q., and T.Y. supervised the project. S.X. wrote the manuscript in consultation with all the authors. All authors discussed the results and commented on the manuscript.

## Competing interests

The authors declare no competing interests.

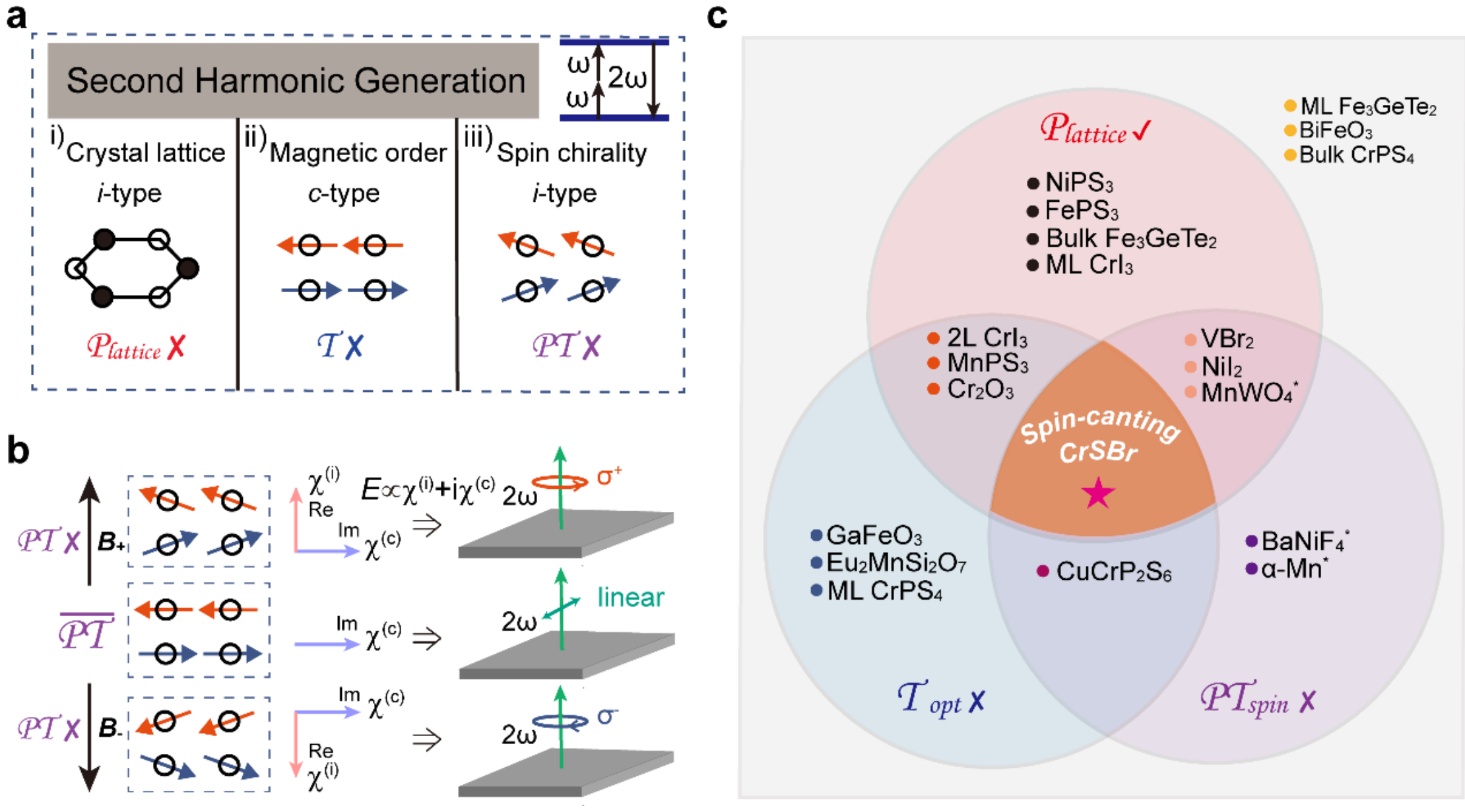

**Figure 1 | Symmetry origins of switchable magnetochirality in centrosymmetric magnets. (a)** Classification of electric-dipole (ED) second-harmonic generation (SHG) origins based on symmetry breaking: **(i)** Crystal lattice-induced *i*-type SHG from broken lattice inversion ($\mathcal{P}_{\text{lattice}}$ ×); **(ii)** Magnetically induced *c*-type SHG originating from $\mathcal{T}$-broken ($\mathcal{T}$×) magnetic order; **(iii)** Spin-chirality-driven *i*-type SHG arising from broken $\mathcal{PT}$ symmetry in non-collinear or canted spin textures ($\mathcal{PT}$×). **(b)** Mechanism of switchable magnetochirality via field-induced $\mathcal{PT}$ symmetry breaking. The collinear ground state preserves $\mathcal{PT}$ symmetry, yielding purely linearly polarized *c*-type SHG. Applying an external magnetic field ($B_{\pm}$) cants the spins, breaking the $\mathcal{PT}$ symmetry and activating the real $\chi^{(i)}$ component. Its coherent interference with the intrinsic imaginary $\chi^{(c)}$ generates switchable circularly polarized emission ($\sigma^+$ or $\sigma^-$). **(c)** Symmetry-based classification of magnetic materials. The Venn diagram defines three prerequisites for achieving chiral interference: a centrosymmetric lattice ($\overline{\mathcal{P}_{lattice}}$), intrinsic *c*-type SHG ($\mathcal{T}_{opt}$×), and chiral *i*-type SHG ($\mathcal{PT}_{spin}$×). Bilayer (2L) CrSBr (star) occupies the exact intersection of all three prerequisites. Asterisks denote symmetry-based candidates for which SHG response has not been experimentally established.

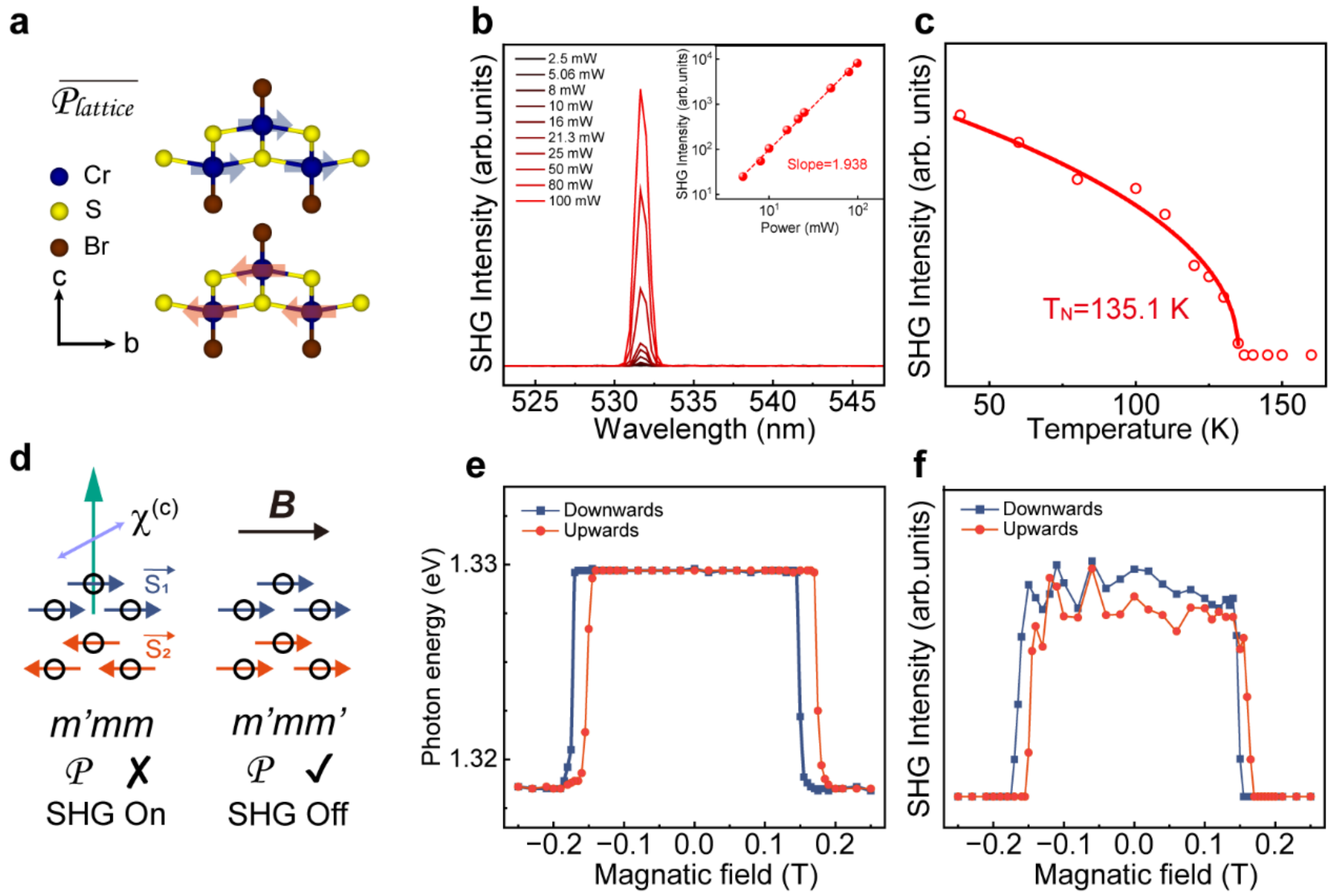

**Figure 2 | Intrinsic symmetry breaking and *c*-type SHG of the antiferromagnetic ground state in 2L CrSBr. (a)** Crystal and magnetic structures of the A-type antiferromagnetic (AFM) ground state in bilayer (2L) CrSBr. **(b)** Power dependence of the SHG intensity at 2 K. The slope of ≈ 1.938 (inset) confirms the second-order nonlinear optical nature. **(c)** Temperature dependence of the SHG intensity. The signal emerges below the Néel temperature ($T_N$ ≈ 135.1 K), establishing its magnetic origin. **(d)** Spin-flip transition and symmetry restoration. The AFM state (magnetic point group $m'mm$) allows c-type SHG due to broken $\mathcal{P}$ symmetry. In contrast, the field-polarized ferromagnetic (FM) state (magnetic point group $m'mm'$) restores $\mathcal{P}$ symmetry, forbidding all ED-SHG. **(e, f)** Magnetic field dependence of **(e)** the exciton photoluminescence (PL) peak energy and **(f)** the SHG intensity across the spin-flip transition. The hysteretic SHG quenching upon entering the FM state in **(f)** experimentally verifies the symmetry restoration predicted in **(d)**.

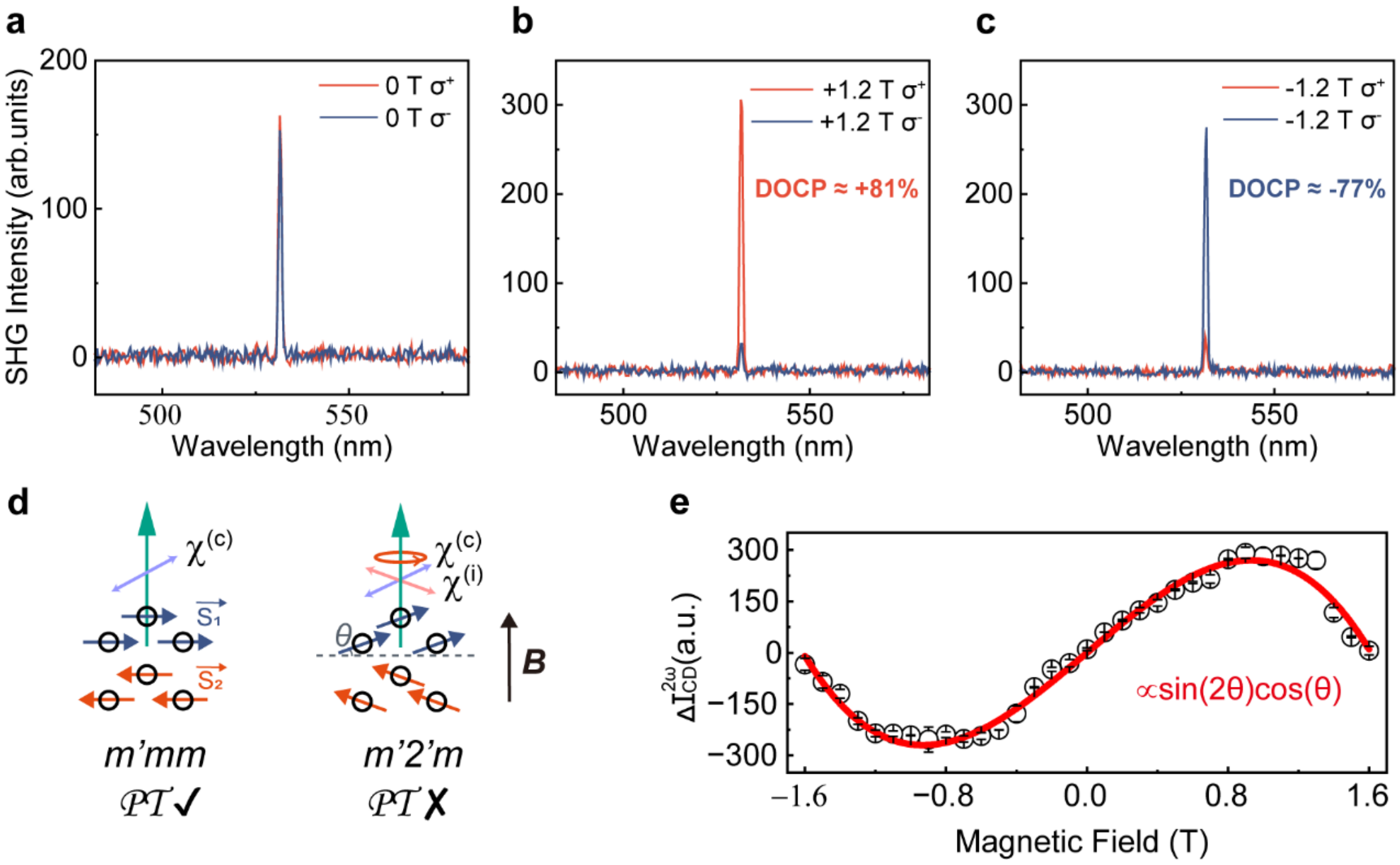

**Figure 3 | Giant magnetically switchable chirality driven by field-induced $\mathcal{PT}$ symmetry breaking. (a–c)** Circularly-resolved SHG spectra at **(a)** 0 T, **(b)** +1.2 T, and **(c)** −1.2 T. The 0 T collinear ground state preserves $\mathcal{PT}$ symmetry, yielding identical left- ($\sigma^+$, red) and right-handed ($\sigma^-$, blue) emission intensities **(a)**. An out-of-plane field of +1.2 T **(b)** induces predominantly $\sigma^+$ emission, yielding a giant SHG circular dichroism (CD) with DOCP ≈ +81%. Reversing the field to −1.2 T **(c)** switches the emission to $\sigma^-$, reversing the DOCP to ≈ -77%. **(d)** Spin configuration and symmetry reduction. A field-induced spin canting angle ($\theta$) reduces the magnetic point group from $m'mm$ to $m'2'm$. This breaks the $\mathcal{PT}$ symmetry and activates an $i$-type susceptibility ($\chi^{(i)}$) that interferes coherently with the intrinsic $c$-type susceptibility($\chi^{(c)}$). **(e)** Verification of the interference model. The SHG CD intensity, $\Delta I_{CD}^{2\omega} = I(\sigma^+) - I(\sigma^-)$, is plotted as a function of the magnetic field. The experimental data (circles) are reproduced by the theoretical function $\Delta I_{CD}^{2\omega} \propto \mathrm{Im}\left[\chi^{(c)} \cdot \chi^{(i)*}\right] \propto sin2\theta \cdot cos\theta$ (red solid line), confirming the coherent interference mechanism.

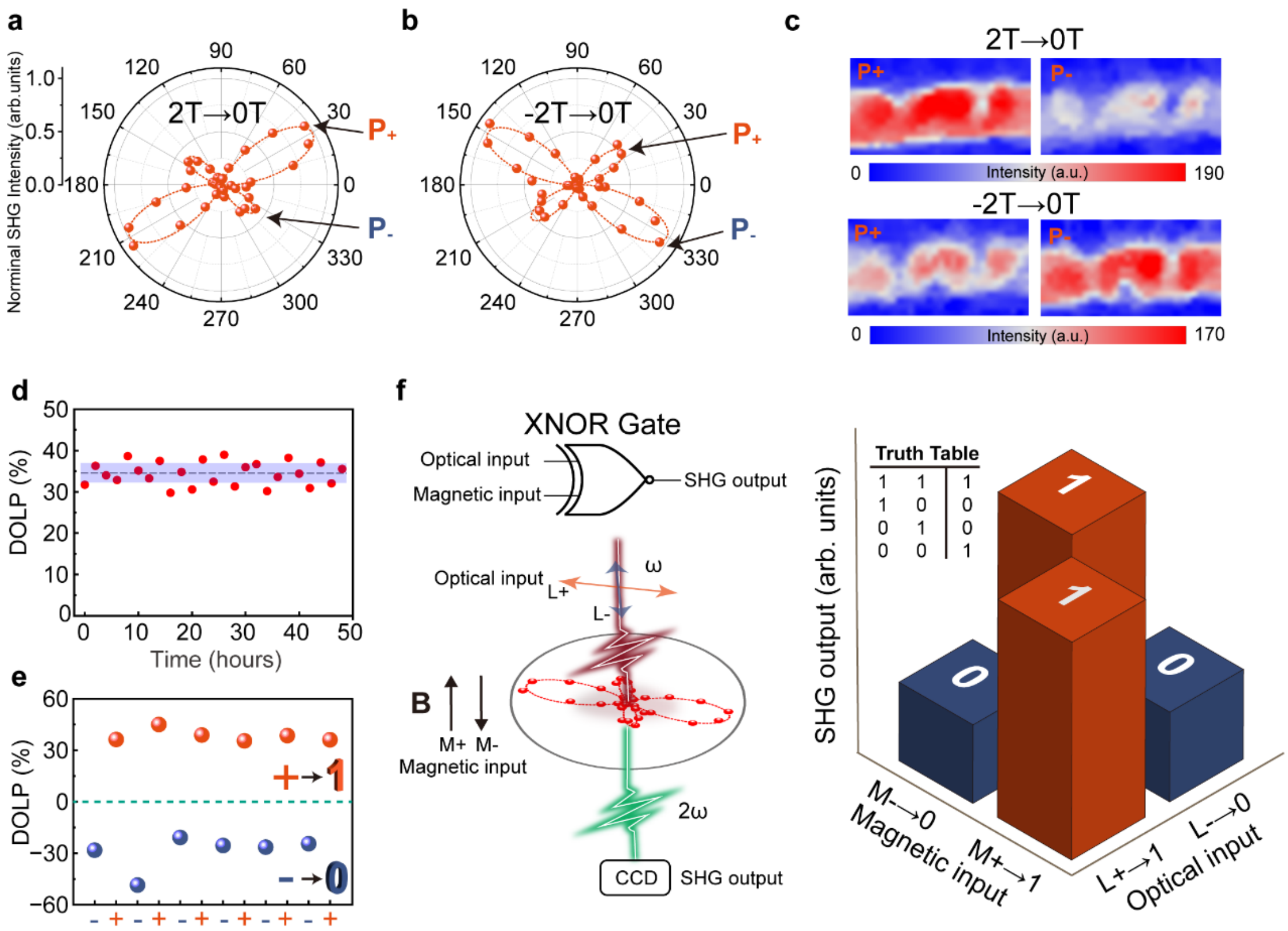

**Figure 4 | Non-volatile magneto-optical memory and XNOR logic operations. (a, b)** Detection of opposite remanent spin chirality. Linearly polarized rotational anisotropy (RA) SHG pattern measured at zero field (0 T) following magnetic saturation at **(a)** +2 T and **(b)** −2 T. The principal polarization axes (labeled $P_+$ and $P_-$) rotate in opposite directions depending on the magnetic history. **(c)** Spatial SHG intensity mapping. The maps are projected onto the $P_+$ and $P_-$ polarization axes for the two remanent states. The SHG intensity is globally dominated by the $P_+$ component across the entire flake in the positive remanent state (2 T→0 T, top row), whereas this contrast reverses ($P_- > P_+$) in the negative remanent state (−2 T→0 T, bottom row). **(d)** Temporal stability of the positive remanent state. The degree of linear polarization (DOLP), monitored at 0 T for 48 hours following magnetic saturation at +2 T, remains nearly constant, demonstrating long-term retention of the remanent optical state. **(e)** Reversible binary encoding. The DOLP serves as the readout signal. The binary states ('1' and '0') are written via out-of-plane magnetic fields with alternating polarities and

read out non-destructively through the optical DOLP, which exhibits deterministic switching between positive and negative values. **(f)** XNOR logic gate operation. The linear polarization axes and the opposite magnetic polarities are assigned as two-state logic inputs. The read-out SHG intensity maps out the XNOR truth table, where the strong and weak signals act as ‘1’ and ‘0’ logic.

**Supplementary Note 1 | Symmetry Analysis and Susceptibility Derivation for Bilayer (2L) CrSBr of Different Magnetic States**

The crystallographic point group of the bilayer CrSBr is $mmm$, which possesses inversion symmetry, thereby forbidding electric-dipole (ED) SHG. Consequently, the ED-SHG susceptibility $\chi^{(i)}_{crystal}$ vanishes identically.

**1. AFM Ground State**

In the A-type antiferromagnetic (AFM) ground state, the magnetic point group is $m'mm$. This symmetry breaks spatial inversion ($\mathcal{P}$) and time-reversal ($\mathcal{T}$) symmetries individually but preserves the combined $\mathcal{PT}$ symmetry. The surviving SHG susceptibility tensor elements are restricted to the $c$-type ($\mathcal{T}$-odd). The non-zero components of the $c$-type susceptibility tensor are given by:

$$\chi^{(c)}_{m'mm}(2\omega) = \begin{pmatrix} \begin{pmatrix} \chi^{xxx} \\ 0 \\ 0 \end{pmatrix} & \begin{pmatrix} 0 \\ \chi^{xyy} \\ 0 \end{pmatrix} & \begin{pmatrix} 0 \\ 0 \\ \chi^{xzz} \end{pmatrix} \\ \begin{pmatrix} 0 \\ \chi^{yxy} \\ 0 \end{pmatrix} & \begin{pmatrix} \chi^{yxy} \\ 0 \\ 0 \end{pmatrix} & \begin{pmatrix} 0 \\ 0 \\ 0 \end{pmatrix} \\ \begin{pmatrix} 0 \\ 0 \\ \chi^{zxz} \end{pmatrix} & \begin{pmatrix} 0 \\ 0 \\ 0 \end{pmatrix} & \begin{pmatrix} \chi^{zxz} \\ 0 \\ 0 \end{pmatrix} \end{pmatrix}$$

Thus, the in-plane components relevant to the normal incidence geometry are $\chi^{xxx}$, $\chi^{xyy}$, $\chi^{yxy}$.

**2. Spin-canting state aligned with $c$ axis**

Under a small out-of-plane magnetic field, the induced spin-canting reduces the symmetry to the $m'2'm$ magnetic point group. This state breaks the $\mathcal{PT}$ symmetry, activating a new set of $i$-type ($\mathcal{T}$-even) tensor elements. The total ED-SHG response includes contributions from both $c$-type and $i$-type components. The $c$-type tensor components for $m'2'm$ remain identical to those of the $m'mm$ group.

$$\chi^{(c)}_{m'2'm}(2\omega) = \begin{pmatrix} \begin{pmatrix} \chi^{xxx} \\ 0 \\ 0 \end{pmatrix} & \begin{pmatrix} 0 \\ \chi^{xyy} \\ 0 \end{pmatrix} & \begin{pmatrix} 0 \\ 0 \\ \chi^{xzz} \end{pmatrix} \\ \begin{pmatrix} 0 \\ \chi^{yxy} \\ 0 \end{pmatrix} & \begin{pmatrix} \chi^{yxy} \\ 0 \\ 0 \end{pmatrix} & \begin{pmatrix} 0 \\ 0 \\ 0 \end{pmatrix} \\ \begin{pmatrix} 0 \\ 0 \\ \chi^{zxz} \end{pmatrix} & \begin{pmatrix} 0 \\ 0 \\ 0 \end{pmatrix} & \begin{pmatrix} \chi^{zxz} \\ 0 \\ 0 \end{pmatrix} \end{pmatrix}$$

The field-activated $i$-type susceptibility tensor for $m'2'm$ is given by:

$$\chi^{(i)}_{m'2'm}(2\omega) = \begin{pmatrix} \begin{pmatrix} 0 \\ \chi^{xxy} \\ 0 \end{pmatrix} & \begin{pmatrix} \chi^{xxy} \\ 0 \\ 0 \end{pmatrix} & \begin{pmatrix} 0 \\ 0 \\ 0 \end{pmatrix} \\ \begin{pmatrix} \chi^{yxx} \\ 0 \\ 0 \end{pmatrix} & \begin{pmatrix} 0 \\ \chi^{yyy} \\ 0 \end{pmatrix} & \begin{pmatrix} 0 \\ 0 \\ \chi^{yzz} \end{pmatrix} \\ \begin{pmatrix} 0 \\ 0 \\ 0 \end{pmatrix} & \begin{pmatrix} 0 \\ 0 \\ \chi^{zyz} \end{pmatrix} & \begin{pmatrix} 0 \\ \chi^{zyz} \\ 0 \end{pmatrix} \end{pmatrix}$$

Thus, in the spin-canting state, the in-plane SHG response is governed by a total of six non-zero tensor components, comprising the three intrinsic $\chi^{(c)}$ elements ($\chi^{xxx}$, $\chi^{xyy}$, $\chi^{yxy}$) and the three spin-chirality-induced $\chi^{(i)}$ elements ($\chi^{xxy}$, $\chi^{yxx}$, $\chi^{yyy}$).

**3. Spin-canting state aligned with *a* axis**

Applying a minute magnetic field along the $a$-axis induces a spin-canting state aligned with the $a$-axis, transforming the magnetic point group to $22'2'$. Although this state breaks $\mathcal{PT}$ symmetry, the in-plane $i$-type tensor components relevant to the normal-incidence geometry remain forbidden by the residual symmetries. Consequently, only $c$-type ED-SHG contributes in the present geometry. The allowed c-type tensor components for $22'2'$ are consistent with those of the $m'mm$ symmetry for in-plane terms:

$$\chi^{(c)}_{22'2'}(2\omega) = \begin{pmatrix} \begin{pmatrix} \chi^{xxx} \\ 0 \\ 0 \end{pmatrix} & \begin{pmatrix} 0 \\ \chi^{xyy} \\ 0 \end{pmatrix} & \begin{pmatrix} 0 \\ 0 \\ \chi^{xzz} \end{pmatrix} \\ \begin{pmatrix} 0 \\ \chi^{yxy} \\ 0 \end{pmatrix} & \begin{pmatrix} \chi^{yxy} \\ 0 \\ 0 \end{pmatrix} & \begin{pmatrix} 0 \\ 0 \\ 0 \end{pmatrix} \\ \begin{pmatrix} 0 \\ 0 \\ \chi^{zxz} \end{pmatrix} & \begin{pmatrix} 0 \\ 0 \\ 0 \end{pmatrix} & \begin{pmatrix} \chi^{zxz} \\ 0 \\ 0 \end{pmatrix} \end{pmatrix}$$

**4. Forced FM state**

When the applied magnetic field is sufficiently strong to align all spins parallel to its direction, the system enters a forced ferromagnetic (FM) state. The magnetic point group transforms to $m'mm'$, which restores the $\mathcal{P}$ symmetry. Consequently, the ED SHG is forbidden.

**Supplementary Note 2 | Theoretical Derivation of SHG Circular Dichroism (CD) in Spin-canting 2L CrSBr**

Based on the susceptibility analysis in the **Supplementary Note 1**, the in-plane SHG response of the spin-canting 2L CrSBr is governed by a total of six non-zero tensor components, comprising the three intrinsic $\chi^{(c)}$ elements ($\chi^{xxx}$, $\chi^{xyy}$, $\chi^{yxy}$) and the three spin-chirality-induced $\chi^{(i)}$ elements ($\chi^{xxy}$, $\chi^{yxx}$, $\chi^{yyy}$). The components of the induced second-harmonic polarization vector $\mathbf{P(2\omega)}$ are expressed as:

$$P_x(2\omega) = \chi^{xxx}E_x^2(\omega) + 2\chi^{xxy}E_x(\omega)E_y(\omega) + \chi^{xyy}E_y^2(\omega)$$

$$P_y(2\omega) = \chi^{yxx}E_x^2(\omega) + 2\chi^{yxy}E_x(\omega)E_y(\omega) + \chi^{yyy}E_y^2(\omega)$$

We consider circularly polarized excitation propagating along the z-axis. The fundamental electric field is defined as $\mathbf{E(\omega)} = \frac{1}{\sqrt{2}}E(\omega)(1, \pm i)$, where the choice of $(1, +i)$ denotes left-hand circular polarization (LCP) and $(1, -i)$ denotes right-hand circular polarization (RCP). Substituting these fields into the polarization equations yields:

$$P_x(2\omega) = \frac{1}{2}(\chi^{xxx} \pm 2i\chi^{xxy} - \chi^{xyy})E^2(\omega)$$
$$P_y(2\omega) = \frac{1}{2}(\chi^{yxx} \pm 2i\chi^{yxy} - \chi^{yyy})E^2(\omega)$$

For the detection path, the generated SHG field is projected onto the circular basis states. The electric field amplitudes projected onto the Left ($\hat{e}_L \propto (1, -i)$) and Right ($\hat{e}_R \propto (1, i)$) circular collection states are given by $E_{out} \propto P_x \mp iP_y$. Here, we denote the polarization-resolved intensities as $I_{\alpha\beta}$, where $\alpha$ and $\beta$ represent the excitation and collection helicities ($L$ or $R$), respectively. For instance, $I_{LL}$ corresponds to LCP excitation and LCP collection. The intensities for the four polarization configurations are derived as follows:

$$I_{LL} \propto \left|P_x - iP_y\right|^2 = |\chi^{xxx} + 2i\chi^{xxy} - \chi^{xyy} - i\chi^{yxx} + 2\chi^{yxy} + i\chi^{yyy}|^2$$

$$I_{LR} \propto \left|P_x + iP_y\right|^2 = |\chi^{xxx} + 2i\chi^{xxy} - \chi^{xyy} + i\chi^{yxx} - 2\chi^{yxy} - i\chi^{yyy}|^2$$

$$I_{RL} \propto \left|P_x - iP_y\right|^2 = |\chi^{xxx} - 2i\chi^{xxy} - \chi^{xyy} - i\chi^{yxx} - 2\chi^{yxy} + i\chi^{yyy}|^2$$

$$I_{RR} \propto \left|P_x + iP_y\right|^2 = |\chi^{xxx} - 2i\chi^{xxy} - \chi^{xyy} + i\chi^{yxx} + 2\chi^{yxy} - i\chi^{yyy}|^2$$

The total collected LCP($I_{\sigma^+}$) and RCP ($I_{\sigma^-}$) intensities are defined as the sum of contributions from both excitations:

$$I_{\sigma^+} = I_{LL} + I_{RL}$$

$$I_{\sigma^-} = I_{LR} + I_{RR}$$

Here, we use the term SHG Circular Dichroism (CD) to denote the emitted-helicity intensity difference. The SHG CD $\Delta I_{CD}^{2\omega} = I_{\sigma^+} - I_{\sigma^-}$. Expanding the intensity terms yields:

$$\Delta I_{CD}^{2\omega} \propto |\chi^{xxx} + 2i\chi^{xxy} - \chi^{xyy} - i\chi^{yxx} + 2\chi^{yxy} + i\chi^{yyy}|^2$$
$$+|\chi^{xxx} - 2i\chi^{xxy} - \chi^{xyy} - i\chi^{yxx} - 2\chi^{yxy} + i\chi^{yyy}|^2$$
$$-|\chi^{xxx} + 2i\chi^{xxy} - \chi^{xyy} + i\chi^{yxx} - 2\chi^{yxy} - i\chi^{yyy}|^2$$
$$-|\chi^{xxx} - 2i\chi^{xxy} - \chi^{xyy} + i\chi^{yxx} + 2\chi^{yxy} - i\chi^{yyy}|^2$$

Without imposing a specific phase relation, all six susceptibility components are generally complex at finite optical frequencies. For clarity, we introduce the effective $c$-type combinations

$$\mathcal{C}_1 \equiv \chi^{xxx} - \chi^{xyy},\ \mathcal{C}_2 \equiv 2\chi^{yxy},$$

and the effective $i$-type combinations

$$\mathcal{I}_1 \equiv 2\chi^{xxy},\ \mathcal{I}_2 \equiv \chi^{yxx} - \chi^{yyy}.$$

The four polarization-resolved SHG intensities can then be written as

$$I_{LL} \propto |\mathcal{C}_1 + \mathcal{C}_2 + i(\mathcal{I}_1 - \mathcal{I}_2)|^2,$$

$$I_{LR} \propto |\mathcal{C}_1 - \mathcal{C}_2 + i(\mathcal{I}_1 + \mathcal{I}_2)|^2,$$

$$I_{RL} \propto |\mathcal{C}_1 - \mathcal{C}_2 - i(\mathcal{I}_1 + \mathcal{I}_2)|^2,$$

and

$$I_{RR} \propto |\mathcal{C}_1 + \mathcal{C}_2 - i(\mathcal{I}_1 - \mathcal{I}_2)|^2.$$

Using

$$|\ X + iY\ |^2 - |\ X - iY\ |^2 = 4\,\mathrm{Im}(XY^*),$$

the emitted-helicity intensity difference becomes

$$\Delta I_{CD}^{2\omega} = I_{LL} + I_{RL} - I_{LR} - I_{RR}$$

$$\propto \mathrm{Im}[\mathcal{C}_2\mathcal{I}_1^* - \mathcal{C}_1\mathcal{I}_2^*]$$

$$\propto \mathrm{Im}[4\chi^{yxy}(\chi^{xxy})^* - (\chi^{xxx} - \chi^{xyy})(\chi^{yxx} - \chi^{yyy})^*]$$

This general $\Delta I_{CD}^{2\omega}$ expression contains exclusively interference terms between the *c*-type and *i*-type susceptibility channels; all terms involving only *c*-type or only *i*-type components cancel identically. Schematically:

$$\Delta I_{CD}^{2\omega} \propto \mathrm{Im}\left[\chi^{(c)} \cdot \chi^{(i)*}\right] \propto sin2\theta \cdot cos\theta$$

Physically, since $\chi^{(c)}$ scales with the Néel vector ($\propto cos\theta$) and $\chi^{(i)}$ scales with the spin chirality ($\propto sin2\theta$), this confirms that the experimentally observed SHG CD arises from the coherent interference between these two distinct order parameters. In the AFM ground state ($B = 0$), the spin chirality vanishes ($\chi^{(i)} \to 0$), leading to $\Delta I_{CD}^{2\omega} = 0$. Conversely, in the spin-canting state ($B \neq 0$), the field-activated $\chi^{(i)}$ switches on this interference channel, resulting in a magnetically controllable CD response.

Equivalently, when considering phase, writing $\mathcal{C}_j = |\mathcal{C}_j| e^{i\phi_{\mathcal{C}_j}}$ and $\mathcal{I}_j = |\mathcal{I}_j| e^{i\phi_{\mathcal{I}_j}}$, we obtain

$$\Delta I_{CD}^{2\omega} \propto |\mathcal{C}_2||\mathcal{I}_1| \sin\left(\phi_{\mathcal{C}_2} - \phi_{\mathcal{I}_1}\right) - |\mathcal{C}_1||\mathcal{I}_2| \sin\left(\phi_{\mathcal{C}_1} - \phi_{\mathcal{I}_2}\right).$$

Thus, the magnitude of the emitted-helicity contrast is governed by both the amplitudes and the relative phases of the two magnetic SHG channels. For fixed susceptibility amplitudes, the magnitude of each interference contribution is maximized when the corresponding *c*-type and *i*-type channels differ in phase by $\pm 90°$, whereas it vanishes when they are in phase or antiphase. In the ideal non-resonant and non-dissipative limit, the purely imaginary c-type and purely real i-type susceptibilities ($\chi^{(i)} \in \mathbb{R}$, $\chi^{(c)} \in i\mathbb{R}$) naturally realize this phase-orthogonal condition.

**Supplementary Note 3 | Theoretical Derivation of Rotational Anisotropy SHG (RA-SHG) Patterns in 2L CrSBr**

The electric field of the linearly polarized fundamental light is given by $\mathbf{E}(\boldsymbol{\omega}) = E(\omega)(cos\varphi, sin\varphi)$. As justified in **Supplementary Note 2**, we obtain the induced polarization components:

$$P_x = \chi^{xxx}\cos^2\varphi + 2\chi^{xxy}\cos\varphi\sin\varphi + \chi^{xyy}\sin^2\varphi$$

$$P_y = \chi^{yxx}\cos^2\varphi + 2\chi^{yxy}\cos\varphi\sin\varphi + \chi^{yyy}\sin^2\varphi$$

The SHG intensity is measured in the parallel and crossed polarization geometries. The projected electric field amplitudes are:

$$E_{\parallel} \propto P_x\cos\varphi + P_y\sin\varphi$$

$$E_{\perp} \propto -P_x\sin\varphi + P_y\cos\varphi$$

For simplicity, the RA-SHG patterns are calculated within a real-valued effective-amplitude approximation. The intensities for the parallel and crossed polarization configurations are derived as follows:

$$I_{\parallel} \propto |\chi^{xxx}\cos^3\varphi + (\chi^{xyy} + 2\chi^{yxy})\sin^2\varphi\cos\varphi + \chi^{yyy}\sin^3\varphi + (\chi^{yxx} + 2\chi^{xxy})\cos^2\varphi\sin\varphi|^2$$

$$I_{\perp} \propto |\chi^{yxx}\cos^3\varphi + (\chi^{yyy} - 2\chi^{xxy})\sin^2\varphi\cos\varphi - \chi^{xyy}\sin^3\varphi + (2\chi^{yxy} - \chi^{xxx})\cos^2\varphi\sin\varphi|^2$$

In the AFM ground state $(\chi^{xxy},\ \chi^{yxx},\ \chi^{yyy} \to 0)$,

$$I_{\parallel} \propto |\chi^{xxx}\cos^3\varphi + (\chi^{xyy} + 2\chi^{yxy})\sin^2\varphi\cos\varphi|^2$$

$$I_{\perp} \propto |-\chi^{xyy}\sin^3\varphi + (2\chi^{yxy} - \chi^{xxx})\cos^2\varphi\sin\varphi|^2$$

Hence, in the collinear AFM state, the absence of the $i$-type susceptibility yields a mirror-symmetric four-lobe RA-SHG pattern with four symmetry-related lobes of equal intensity. A finite $i$-type susceptibility introduces a sign-sensitive interference term with the intrinsic $c$-type response, producing two major and two minor lobes; reversing the sign of $\chi^{(i)}$ interchanges these lobes and results in oppositely distorted RA-SHG patterns.

| Magnetic State | Magnetic Point Group | $\mathcal{PT}$ Symmetry | $\mathcal{P}$ Symmetry | i-type (Chiral) Susceptibility | c-type Susceptibility | Macroscopic SHG |
|---|---|---|---|---|---|---|
| **Collinear AFM** (Ground state, $B = 0$) | $m'mm$ | Preserved | Broken | Forbidden | Allowed | Linearly Polarized |
| **Collinear FM** (Saturated state, $B > B_c$) | $m'mm'$ | Broken ($\kappa$=0) | Preserved | Forbidden | Forbidden | Forbidden |
| **Spin-Canting AFM** (**aligned with *c* axis**) | $m'2'm$ | Broken | Broken | Allowed | Allowed | Circularly Polarized |
| **Spin-Canting AFM** (**aligned with *a* axis**) | $22'2'$ | Broken | Broken | Forbidden | Allowed | Linearly Polarized |

**Supplementary Table S1 | Group-theoretical classification, symmetry operations, and non-vanishing SHG susceptibilities for 2L CrSBr across distinct magnetic phases.** The evolution of the magnetic point group dictates the allowed SHG susceptibility tensors: the unperturbed collinear AFM ground state permits only $\chi^{(c)}$, while the spin-canting state breaks $\mathcal{PT}$ symmetry, activating the chiral $\chi^{(i)}$ alongside $\chi^{(c)}$ to generate circular polarization. In contrast, the saturated collinear FM state restores $\mathcal{P}$ symmetry, quenching all electric-dipole SHG responses. This deterministic symmetry evolution under external fields provides the symmetry basis for realizing and manipulating optical magnetochirality in 2L CrSBr.

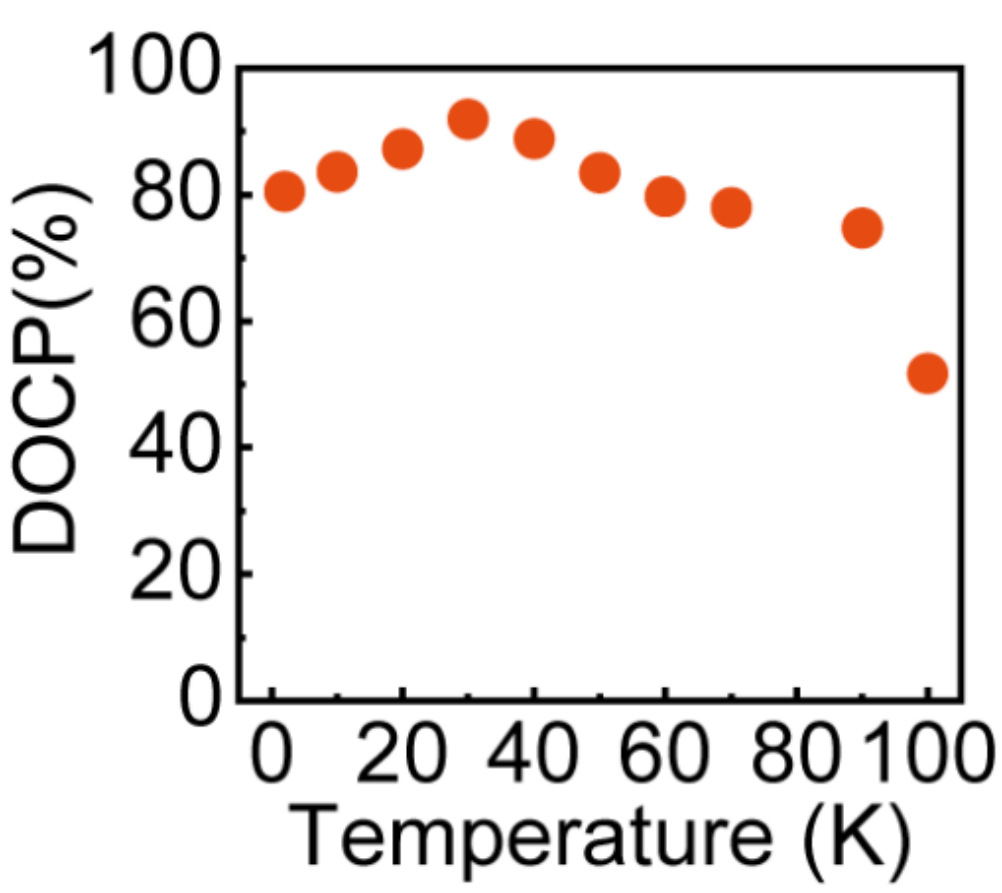

**Supplementary Figure S1 | Temperature dependence of the degree of circular polarization (DOCP).** Evolution of the SHG DOCP as a function of temperature under an appropriate out-of-plane magnetic field. The giant nonlinear optical magnetochirality is robustly maintained to 90 K, reaching approximately 90% near 30 K. As the temperature further increases towards the Néel temperature, thermal fluctuations gradually disrupt the delicate spin-canting configuration and the underlying long-range magnetic order, leading to a decrease in the macroscopic chiral SHG response.

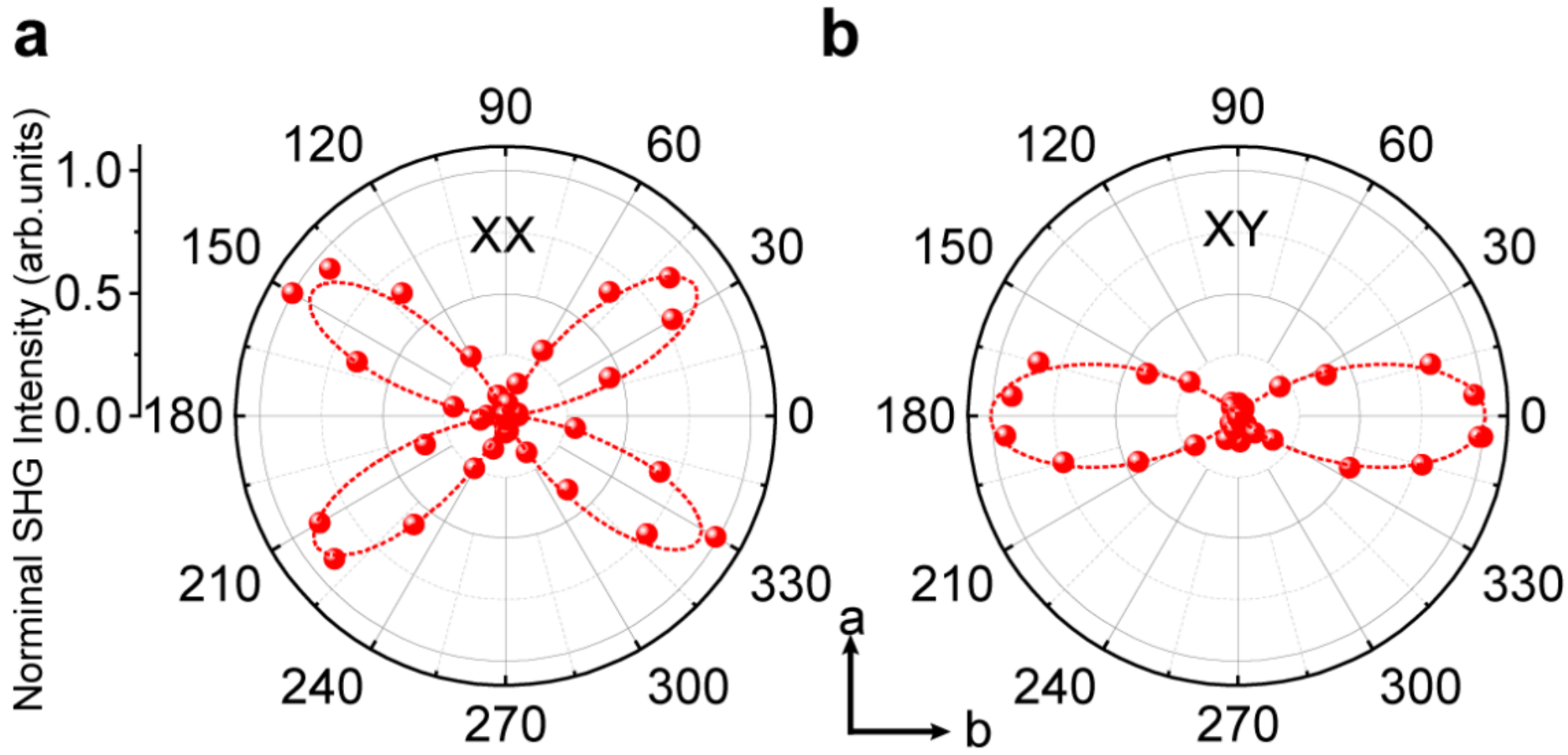

**Supplementary Figure S2 | Symmetrical rotational anisotropy (RA) SHG patterns of the unperturbed collinear AFM ground state. (a)** RA-SHG intensity under parallel (XX) polarization. The geometrically symmetric four-lobe pattern is the hallmark of the collinear A-type AFM ground state. In this $\mathcal{PT}$-preserved phase, the chiral $i$-type susceptibility is forbidden, and the nonlinear optical response is exclusively governed by the intrinsic $\chi^{(c)}$. **(b)** The corresponding RA-SHG pattern under cross (XY) polarization, exhibiting a highly symmetric two-lobe structure aligned with the crystallographic $b$ axis. The red circles represent data points, and the dashed lines are theoretical fits derived from the $m'mm$ magnetic point group. These symmetric baseline patterns contrast with the distorted RA-SHG profiles observed in the remanent canting states (Fig. 4a), validating that the principal axis rotation in the remanent state is driven by the emergent $\chi^{(i)}$ susceptibility.

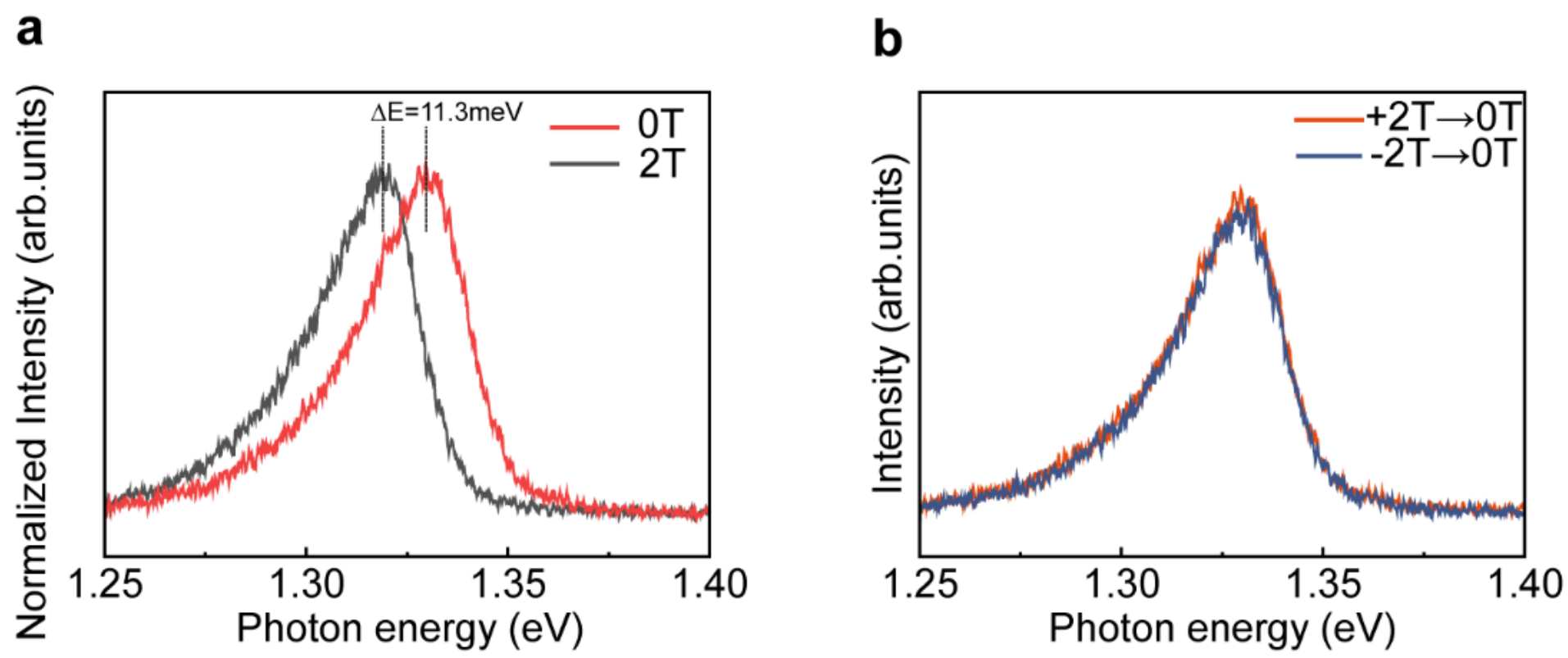

**Supplementary Figure S3 | Insensitivity of conventional linear optical probes to remanent spin-canting states. (a)** Exciton photoluminescence (PL) spectra of 2L CrSBr at 0 T (AFM ground state) and 2 T (FM state). The transition from the interlayer AFM to the FM configuration induces a characteristic redshift of the exciton energy by $\Delta E = 11.3$ meV, driven by the strong spin-layer coupling. **(b)** Exciton PL spectra measured at 0 T after magnetic saturation at +2 T (red) and -2 T (blue). The spectra of the positive and negative remanent states overlap without any resolvable energy splitting or intensity variation. The absence of a resolvable PL difference is consistent with the small magnitude of the inferred remanent spin canting.

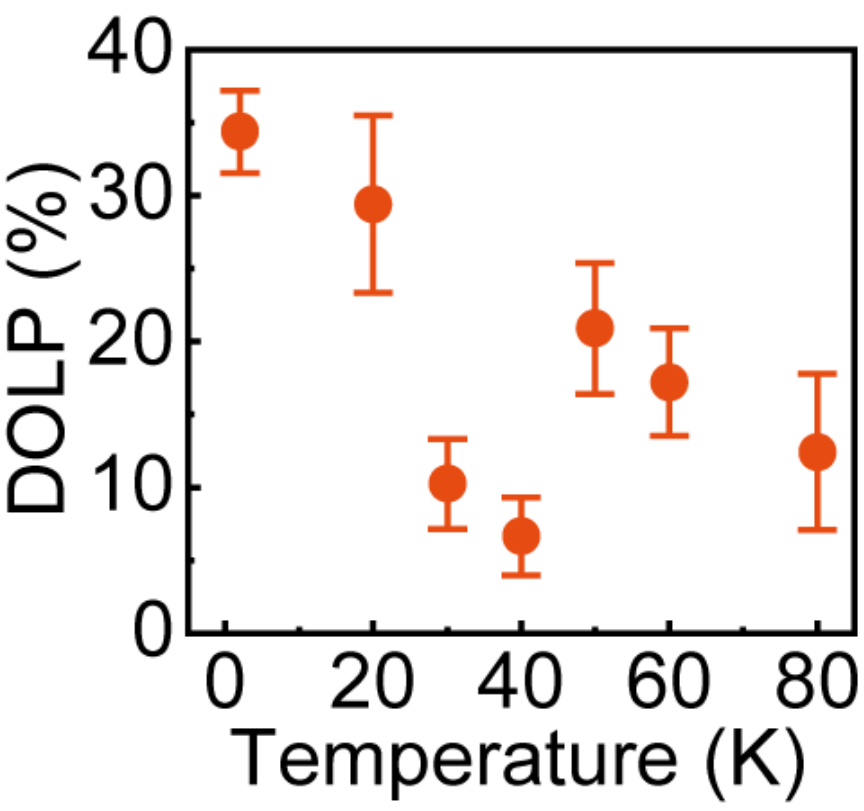

**Supplementary Figure S4 | Temperature dependence of the degree of linear polarization (DOLP).** The DOLP exhibits a maximum of approximately 35% at base temperature but undergoes a rapid decay as thermal fluctuations increase, effectively quenching to a minimum around 30–40 K, indicating a characteristic magnetic transition near 40 K that corroborates the local ferromagnetic pinning of the frustrated spin configuration in its metastable canting state.